\documentclass[12pt,a4paper]{article}
\usepackage{amsmath}
\usepackage{amssymb}
\usepackage{graphicx}

\begin{document}

\title{\bf On Unparticles and $K^+\rightarrow \pi^+ +{\rm Missing~Energy}$}
\author{\vspace{0.8cm}\\Yunfei Wu\ \ and\ \ Da-Xin Zhang\thanks{dxzhang@mail.phy.pku.edu.cn}\\\small{
School of Physics, Peking University,  Beijing 100871, China}}
\date{}
\begin{titlepage}
\maketitle \thispagestyle{empty} \vspace{1.5cm}
\begin{abstract}
We analyze the branching ratio and spectrum for the decay mode $K^+
\rightarrow\pi^++{\not}E$(missing energy) in the  unparticle model,
where an unparticle can also serve as the missing energy. A vector
unparticle can even mediate the $K^+ \rightarrow \pi^+ +\nu
\bar{\nu}$ , resulting complicated interference with the Standard
Model.
\end{abstract}


\end{titlepage}

\section{}
The rare decay $K^+ \rightarrow \pi^+ +\nu \bar{\nu}$ is one of the
cleanest decay modes in the Stand Model (SM)\cite{QCD}.  Due to its
smallness within the SM, it might be very sensitive to the new
physics beyond the SM. Since only the $\pi^+$ in the final state
will be detected, the neutrino-anti-neutrino pair will behave as
missing energy. Consequently, in the presence of new physics, the
decay mode $K^+ \rightarrow \pi^+ +\nu \bar{\nu}$ is not only
modified by the new interactions, but also polluted by possible new
final state if it also behaves as  missing energy.

In the Unparticle Model suggested by Georgi\cite{unpar}, an
interesting observation is that a nontrivial scale invariant sector
of scale dimension $d_\mathcal{U}$ might manifest itself at the low
energy as a non-integral number $d_\mathcal{U}$ of invisible
massless particles, dubbed unparticle $\mathcal{U}$.
%It may give
%rise to peculiar missing energy distributions in the process $K^+
%\rightarrow \pi^+ +\nu \bar{\nu}$ through the coupling involving
%both SM fields and unparticle Banks-Zaks fields\cite{BZ}.
In the effective theory below the scale $\Lambda_\mathcal{U}$, the
Banks-Zaks operators\cite{BZ} match onto the unparticles operators,
and the interactions match onto the form\cite{unpar}
\begin{equation}
\frac{\mathcal{C}_\mathcal{U}\Lambda_\mathcal{U}^{d_\mathcal{BZ}-d_\mathcal{U}}}{M_\mathcal{U}^k}\mathcal{O}_{SM}\mathcal{O}_\mathcal{U},
\end{equation}
where $\mathcal {C}_\mathcal{U}$ is a coefficient function. If
$M_\mathcal{U}$ is large enough, the unparticle stuff doesn't couple
strongly to the ordinary particles. Many forms of interactions have
been introduced in the literature, resulting very different features
from those in the SM\cite{unparticle-propagator,CKY,Luo-Zhu,
Chen-Geng, Ding-Yan, Liao-1, Aliev-Cornell-Gaur, Catterall-Sannio,
Li-Wei, Lu-Wang-Wang, Fox-Rajaraman-Shirman, Greiner, Davoudiasl,
Choudhury-Ghosh-Mamta, Chen-He, Mathews-Ravindran, Zhou, Liao-Liu,
minimal-walking, Bander-Feng-Rajaraman-Shirman, Rizzo, ungravity,
Chen-He-Tsai, Zwicky, Kikuchi-Okada, Mohanta-Giri, Huang-Wu, Lenz,
Choudhury-Ghosh, Zhang-Li-Li, Nakayama, Desh-He-Jiang, BGHNS,
Delgado-Espinosa-Quiros, colored-unparticle, Neubert,
Hannestad-Raffelt-Wong, HMS, Desh-Hsu-Jiang, Kumar-Das, BCG, Liao-2,
Majumdar, Alan-Pak-Senol, Freitas-Wyler, GOS, Hur-Ko-Wu,
Anchordoqui-Goldberg, Majhi, McDonald, KMRT, DMR, Kobakhidze,
Balantekin-Ozansoy, Aliev-Savci, LKQWL, Iltan, CHLTW, Lewis, AKTVV,
CHHL, Sahin-Sahin, Stephanov, Krasnikov, HEIDI, Ryttov-Sannino,
JPLee}.

In the present work we will study $K^+\rightarrow \pi^+ +{\rm
Missing~Energy}$ in the Unparticle Model. In this model, the mode
$K^+ \rightarrow \pi^+ +\nu \bar{\nu}$ is modified by the unparticle
mediation, and $K^+ \rightarrow \pi^+ +\mathcal{U}$ also behaves as
the missing energy. They are constrained by the data \cite{pdg}. We
will analyse the spectra of $\pi^+$ in the final states, present
numerical results and give further discussions.

\section{}

In the  Unparticle Model, we study the mode $K^+ \rightarrow \pi^+
+\mathcal{U}$ firstly. The quark-unparticle couplings are taken to
be
\begin{equation}
(\bar qq)_{V\pm
A}\mathcal{O}_\mathcal{U}^{\;\mu}\,,\quad\mbox{and}\quad\,(\bar
qq)_{V\pm A}\partial^{\;\mu}\mathcal{O}_\mathcal{U}\,,\label{sv}
\end{equation}
where $V\pm A$ here refers to $\gamma_\mu(1\pm\gamma_5)$ and the
couplings are omitted. The propagator for the vector unparticles is

\begin{equation}
\Delta_{\mathcal{V}}=-i\int
e^{iPx}\langle0|T(\mathcal{O}_\mathcal{U}^\mu\mathcal{O}_\mathcal{U}^\nu)|0\rangle\;
d^4x=-i\frac{A_{d_{\mathcal{U}}}}{2}\frac{-g^{\mu\nu}+P^\mu
P^\nu/P^2}{\sin(d_\mathcal{U}\pi)}(-P^2-i\epsilon)^{d_\mathcal{U}-2}\,.
\end{equation}
$A_{d_{\mathcal{U}}}$ is defined as\cite{unpar}
\begin{equation}
A_{d_{\mathcal{U}}}=\frac{16\pi^{5/2}}{(2\pi)^{2d_\mathcal{U}}}\frac{\Gamma(d_\mathcal{U}+1/2)}{\Gamma(d_\mathcal{U}-1)\Gamma(2d_\mathcal{U})},
\end{equation}
where $d_\mathcal{U}$ is a non-integral number, counting for the
non-integral number of massless paricle behavior of the
unparticle\cite{unpar}. The effective Hamiltonian for an unparticle
emission process is
\begin{equation}
\mathcal {H}_{eff}^\mathcal{S}=\frac{\mathcal {C}_\mathcal
{S}^q\Lambda_\mathcal {U}^{k-d_\mathcal {U}}}{M_\mathcal {U}^k}(\bar
sd)_{V-A}\partial^{\,\mu}\mathcal {O}_\mathcal {U}
\end{equation}
for the scalar unparticle, and
\begin{equation}
\mathcal {H}_{eff}^\mathcal{V}=\frac{\mathcal {C}_\mathcal
{V}^q\Lambda_\mathcal {U}^{k+1-d_\mathcal {U}}}{M_\mathcal
{U}^k}(\bar sd)_{V-A}\mathcal {O}_\mathcal{U}^{\,\mu}
\end{equation}
for the vector unparticle. We have defined the two dimensional
coefficients corresponding to scalar and vector unparticles
%\footnote{We did this
%just for simplicity. It is important to note the difference with
%literatures\cite{Collider,B} }:
\begin{equation}
\textit{c}_\mathcal{S}^{\,q}=\frac{\mathcal {C}_\mathcal
{S}\Lambda_\mathcal {U}^{k-d_\mathcal {U}}}{M_\mathcal {U}^k}\,,
\quad and \quad \,\textit{c}_\mathcal{V}^{\,q}=\frac{\mathcal
{C}_\mathcal {V}\Lambda_\mathcal {U}^{k+1-d_\mathcal
{U}}}{M_\mathcal {U}^k}\,.\label{cq}
\end{equation}
We  get the hadronic  amplitudes
\begin{equation}
\mathcal{A}_{eff}^S=\textit{c}_\mathcal{S}^{\,q}
\left(f_+(q^2)(k+p)_\mu+f_-(q^2)(k-p)_\mu\right)\partial^{\,\mu}\mathcal
{O}_\mathcal {U},
\end{equation}
\begin{equation}
\mathcal{A}_{eff}^\mathcal{V}=\textit{c}_\mathcal{V}^{\,q}
\left(f_+(q^2)(k+p)_\mu+f_-(q^2)(k-p)_\mu\right)\mathcal
{O}_\mathcal {U}^{\,\mu}.
\end{equation}
The differential width for the scalar and vector unparticles
respectively are,
\begin{eqnarray}
\frac{d\Gamma^\mathcal{SU}}{dE_\pi}&=&\frac{{\textit{c}_\mathcal{S}^{\,q}}^2A_{d_\mathcal{U}}}{4\pi^2m_K}\sqrt{E_\pi^2-m_\pi^2}\left(m_K^2+m_\pi^2-2m_KE_\pi\right)^{d_\mathcal{U}-2}\nonumber\\
&&\times\left[f_+(m_K^2-m_\pi^2)+f_-(m_K^2+m_\pi^2-2m_\pi
E_\pi)\right]^2,
\end{eqnarray}
\begin{eqnarray}
\frac{d\Gamma^\mathcal{VU}}{dE_\pi}&=&\frac{{\textit{c}_\mathcal{V}^{\,q}}^2A_{d_\mathcal{U}}}{4\pi^2m_K}\sqrt{E_\pi^2-m_\pi^2}\left(m_K^2+m_\pi^2-2m_KE_\pi\right)^{d_\mathcal{U}-2}\nonumber\\
&&\times\bigg[f_+^2(m_K^2+m_\pi^2+2m_KE_\pi)+f_-^2(m_K^2+m_\pi^2-2m_KE_\pi)\\
&&+2f_+f_-(m_K^2-m_\pi^2)-(f_+^2+f_-^2+2f_+f_-)\frac{(m_K^2-m_\pi^2)^2}{m_K^2+m_\pi^2-2m_KE_\pi}\bigg]\nonumber.
\end{eqnarray}
%The two formulas above look quite odds for the additional $A_{d_\mathcal{U}}$.

\section{}
In the Unparticle Model, the decay $K^+ \rightarrow \pi^+ +\nu
\bar{\nu}$ receives two sources of contributions. One is from the SM
and the other is from the unparticle mediation. In the SM, the
relevant effective Hamiltonian is\cite{QCD}
\begin{equation}
\mathcal {H}_{eff}=\frac{G_F}{\sqrt{2}} \frac{\alpha}{2\pi
\sin^2\theta_W}\sum\limits_{l=e,\mu,\tau}\left[\,V_{cs}^*V_{cd}X_{NL}^l+V_{ts}^*V_{td}X(x_t)\right](\bar
sd)_{V-A}(\bar\nu_l\nu_l)_{V-A}.\label{h1}
\end{equation}
The index $l=e,\mu,\tau$ denotes the lepton flavor. We have taken
the functions $X$,  $X_{NL}^l$ and the coefficients following
\cite{QCD,QCD2}. The hadronic amplitude is
\begin{eqnarray}
\mathcal{A}_{eff}^{SM}&=&\frac{G_F}{\sqrt{2}} \frac{\alpha}{2\pi
\sin^2\theta_W}\sum\limits_{l=e,\mu,\tau}\left[\,V_{cs}^*V_{cd}X_{NL}^l+V_{ts}^*V_{td}X(x_t)\right]\nonumber\\
&&\times\left(f_+(q^2)(k+p)_\mu+f_-(q^2)(k-p)_\mu\right)(\bar\nu_l\nu_l)_{V-A},
\end{eqnarray}
where $q^2=(k-p)^2$,and $f_\pm (q^2)$ are the form
factors\cite{form2,form}. The differential decay width, where
$E_\pi$ is the pion energy in the rest frame of the decaying kaon,
reads
\begin{eqnarray}
\frac{d\Gamma^{SM}}{dE_\pi}&=&\frac{G_F^2}{4}\frac{\alpha^2\lambda}{(2\pi)^5\sin^4\theta_WM_K^2}\sum\limits_{l=e,
\mu,
\tau}\left[\,V_{cs}^*V_{cd}X_{NL}^l+V_{ts}^*V_{td}X(x_t)\right]^2\nonumber\\
&&\times  f_+^2\bigg[(m_K^2-m_\pi^2-q^2)^2-2m_\pi^2q^2
-\frac{2}{q^2}\bigg(\frac{\lambda^2}{3}+m_\pi^2q^2\bigg)\bigg],\nonumber
\end{eqnarray}
where \begin{equation} m_\pi\leq E_\pi\leq (m_K^2+m_\pi^2)/2m_K,
\end{equation}
\begin{equation}
\lambda=[(m_K^2+m_\pi^2-q^2)^2-4m_k^2m_\pi^2]^{1/2}.
\end{equation}
In the presence of the unparticle, the vector unparticle can also
mediate $K^+ \rightarrow \pi^+ +\nu \bar{\nu}$ if we introduce the
neutrino-unparticle couplings analogue to the quark-unparticle
couplings of (\ref{sv}). This couplings may conserve flavor,
\begin{equation}
\textit{c}_\mathcal{V}^{\,l}(\bar\nu_l\nu_l)_{V-A}\mathcal{O}_\mathcal{U}^{\,\mu}\,,\label{cl}
\end{equation}
the effective interactions of the vector unparticle mediation can be
written as
\begin{equation}
\triangle\mathcal
{H}_{eff}^\mathcal{V}=\textit{c}_\mathcal{V}^{\,q}\textit{c}_\mathcal{V}^{\,l}\frac{A_{d_\mathcal{U}}(-p_\mathcal{U}^2)^{d_\mathcal{U}-2}}{2\sin(d_\mathcal{U}\pi)}
\sum\limits_{l=e,\mu,\tau}(\bar
sd)_{V-A}(\bar\nu_l\nu_l)_{V-A}\,.\label{h2}
\end{equation}
The more general case that the lepton numbers are also violated are
too complicated to be considered here. It is easy to see that the
scalar unparticle couplings
\begin{equation}
\textit{c}_\mathcal{S}^{\,l}(\bar\nu_l\nu_l)_{V-A}\partial^\mu\mathcal{O}_\mathcal{U}\,\nonumber
\end{equation}
do not contribute, as the neutrinos are taken as massless.

In this way, we have more four-fermion interactions which are
different from the SM four-fermion interactions by the dependence on
the momentum transfer. It also has a imaginary part because of the
$(-1)^{d_\mathcal{U}}$, which could induce CP
violation\cite{Chen-Geng}.
%$\triangle\mathcal {H}_{eff}^\mathcal{V}$ look much
%alike except for $d_\mathcal{U}-$2. This small difference is not
%trivial at all.
At the hadronic level we have the amplitude modified as
\begin{eqnarray}
\triangle\mathcal{A}_{eff}^\mathcal{V}&=&\textit{c}_\mathcal{V}^{\,q}\textit{c}_\mathcal{V}^{\,l}\frac{A_{d_\mathcal{U}}(-p_\mathcal{U}^2)^{d_\mathcal{U}-2}}{2\sin(d_\mathcal{U}\pi)}
\sum\limits_{l=e,\mu,\tau}
\left(f_+(q^2)(k+p)_\mu+f_-(q^2)(k-p)_\mu\right)(\bar\nu_l\nu_l)_{V-A}
.
\end{eqnarray}
Here, $p_\mathcal{U}^2=q^2$.

The total differential width for $K^+ \rightarrow \pi^+ +\nu
\bar{\nu}$ including the SM and the vector unparticle contributions
is
\begin{eqnarray}
\frac{d\Gamma^{\mathcal{V}\nu\bar\nu}}{dE_\pi}&=&\frac{\lambda}{2(2\pi)^3M_K^2}\sum\limits_{l=e,
\mu,
\tau}\Bigg|\bigg\{\frac{G_F}{\sqrt2}\frac{\alpha}{2\pi\sin^2\theta_W}\left[\,V_{cs}^*V_{cd}X_{NL}^l+V_{ts}^*V_{td}X(x_t)\right]\nonumber\\
&&+\left(\frac{\textit{c}_\mathcal{V}^{\,q}\textit{c}_\mathcal{V}^{\,l}A_{d_\mathcal{U}}(-q^2)^{d_\mathcal{U}-2}}{2\sin(d_\mathcal{U}\pi)}\right)\bigg\}\Bigg|^2\nonumber\\
&&\times f_+^2\bigg\{(m_K^2-m_\pi^2-q^2)^2-2q^2m_\pi^2
-\frac{2}{q^2}\bigg(\frac{\lambda^2}{3}+m_\pi^2q^2 \bigg)\bigg\}
.\label{vvv}
\end{eqnarray}
%where we neglect the $K^+ \rightarrow \pi^+ +\mathcal{U}$ process.
We take the parameters used in (\ref{h1})
as\cite{QCD,pdg,QCD3,Charm}
\begin{equation}
V_{cs}^* V_{cd}=-0.22006_{-0.00091}^{+0.00093},\quad
V_{td}^*V_{ts}=(-3.13^{+0.20}_{-0.17}+\text
i1.407^{+0.096}_{-0.098})\times10^{-4},\quad
|V_{us}|=0.2248\nonumber.
\end{equation}
\begin{equation}
P_c=\frac{1}{|V_{us}|^4}\left[\frac{2}{3}X_{NL}^e+\frac{1}{3}X_{NL}^\tau\right]=0.375\pm0.024,\quad
X(x_t) = 1.464\pm0.041.,\nonumber
\end{equation}
\begin{equation}
\alpha=1/129,\quad \cos(\theta_W)=0.8817
\end{equation}
We take the formfactors as\cite{pdg}
\begin{equation}
f_+(q^2)=f_+(0)[1+\lambda(q^2/m_{\pi}^2)], \quad f_-(q^2)=-0.332,
\end{equation}
where $\lambda=2.96\times10^{-2}$and take $f_+(0)=0.57$\cite{form2}.
We assume this choice of form factors to be valid for the unparticle
processes. To the next-to-next-to leading logarithm approximation
\cite{QCD3} in the SM,
\begin{equation}
B_{SM}(K^+ \rightarrow \pi^+ +\nu
\bar{\nu})=(8.0\pm1.1)\times10^{-11}.
\end{equation}

\section{}
The process $K^+\rightarrow \pi^+ +{\rm Missing~Energy}$ may contain
$\nu \bar{\nu}$ or $ \mathcal {U}$  in the final state. The decay
width is
\begin{equation}
\frac{d\Gamma(K^+\rightarrow\pi^+{\not}E)}{dE_\pi}=\frac{d\Gamma^{\;\nu\bar\nu}}{dE_\pi}+\frac{d\Gamma^\mathcal{U}}{dE_\pi},\label{dga}
\end{equation}
where ${d\Gamma^{\;\nu\bar\nu}}/{dE_\pi}$ contains the SM and the
(vector) unparticle contributions of (\ref{vvv}). The decay width
(\ref{dga}) is constrained by the experiments.  The E787 and E949
Collaborations at Brookhaven give\cite{experiment}
\begin{equation}
B_{exp}(K^+\rightarrow\pi^+{\not}E)=(14.7_{-8.9}^{+13.0})\times
10^{-11}.
\end{equation}

We now comment on the couplings of the unparticles in (\ref{cq}) and
in (\ref{cl}). On the one hand, the couplings $c^q$'s in (\ref{cq})
are flavor changing while $c^l$'s in (\ref{cl}) are flavor
conserving. If the Unparticle Model followed a GIM-like
mechanism\cite{gim} of the SM for the flavor changing neutral
interactions, the $c^l$'s in (\ref{cl}) could be much larger than
the $c^q$'s in (\ref{cq}). Even if the $c^l$'s are much smaller than
the gauge couplings in the SM, the effect of interference between
the amplitudes of the SM and the vector unparticle mediation might
be comparable to the $\Gamma^\mathcal{U}$.

On the other hand, in case that the couplings $c^l$'s are not much
larger than the $c^q$'s, the contribution of the vector unparticle
mediation is negligible in the process $K^+\rightarrow \pi^+ +{\rm
Missing~Energy}$. The contribution $\Gamma^\mathcal{U}$ can even
dominant over $\Gamma^{\;\nu\bar\nu}$ in the extrame case.

\subsection{}
In this part, we study the effects of the scalar unparticle  in
$K^+\rightarrow\pi^++\nu\bar\nu$ and
$K^+\rightarrow\pi^++\mathcal{U}$. There is no mediation effect for
the scalar unparticle in $K^+\rightarrow\pi^++\nu\bar\nu$. The
scalar unparticle acts only as Missing Energy in the final state. We
plot in Fig. 1 the dependence of the branching ratio
$K^+\rightarrow\pi^+{\not E}$ on the $d_\mathcal{U}$, and in Fig. 2
the dependence on the coupling $\textit{c}_\mathcal{S}^{\,q}$.

\begin{figure}[h]
\raisebox{50pt}{\rotatebox{90}{\footnotesize $B\,(K^+ \rightarrow
\pi^+ +{\not}E)$}}
\includegraphics[width=300pt]{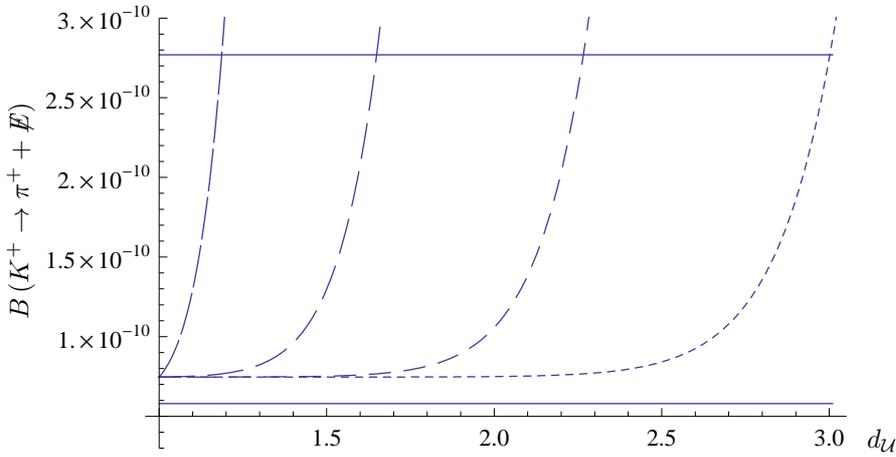}
\makebox(15,0)[br]{\footnotesize$d_\mathcal{U}$}
\caption{\footnotesize Branching ratio versus $d_\mathcal{U}$ for
$K^+ \rightarrow \pi^+ +{\not}E$ in the scalar unparticle model,
with $\textit{c}_\mathcal{S}^{\,q}=1\times 10^{-17}, 1\times
10^{-16}, 1\times 10^{-15}  \text{and} \ 1\times10^{-14}$. The dash
length increases with $\textit{c}_\mathcal{S}^{\,q}$. The horizontal
lines represent the experimental bounds.}
\end{figure}
%It is easy to see the unparticle stuff only takes the place of $(\nu
%\bar\nu)_{V-A}$ in the effective Hamiltonian of $K^+ \rightarrow
%\pi^+ +\mathcal {U}$. The unparticle stuffs behave normally as
%common field except for its non-integral dimensions.

\begin{figure}[h]
\raisebox{60pt}{\rotatebox{90}{\footnotesize $B\,(K^+ \rightarrow
\pi^+ +{\not}E)$}}
\includegraphics[width=300pt]{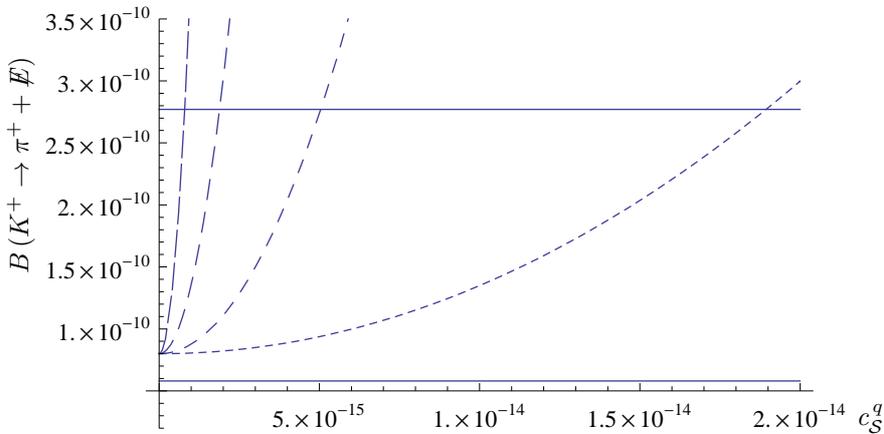}
\makebox(10,0)[br]{\footnotesize$\textit{c}_\mathcal{S}^{\,q}$}
\caption{\footnotesize Branching ratio versus
$\textit{c}_\mathcal{S}^{\,q}$ $K^+ \rightarrow \pi^+ +{\not}E$ in
the scalar unparticle model, where $d_\mathcal{U}$=1.1, 1.3, 1.5,
and 1.7. The dash-length increases with $d_\mathcal{U}$.}
\end{figure}

We find that, if  $d_\mathcal{U} = 1$, no contribution from the
unparticle will reveal. The unparticle contribution goes up with
$d_\mathcal{U}$ and will dominate over the SM contribution. To
fulfill the data,  the scalar unparticle can have a large coupling
$c_\mathcal{S}^{\,q}$ for a small $d_\mathcal{U}$, while the
coupling is constrained for a large $d_\mathcal{U}$.

\begin{figure}[h]
\makebox(20,160)[tr]{$\frac{d\Gamma^{{\not}E}}{dE_\pi}$}
\includegraphics[width=280pt]{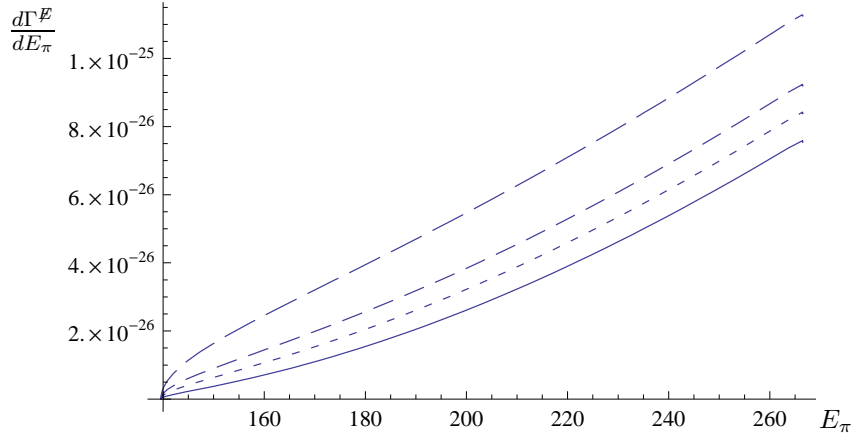}
\makebox(10,0)[br]{\footnotesize$E_\pi$} \par \caption
{\footnotesize {The energy spectrum for the charged $pi$ in the
scalar unparticle model, with $d_\mathcal{U}$ = 1.5,
$\textit{c}_\mathcal{S}^{\,q} = 1\times10^{-15}$ with dash. The
solid line represents the SM result.}}
\end{figure}

We also plot in Fig. 3 the energy spectrum of the charged $\pi$.
Note that apart from the soft $\pi$ region, the energy spectrum is
much like the energy spectrum in the SM with a different number of
neutrino spices.

\subsection{}
The vector unparticle not only acts as the Missing Energy in the
final state, but also mediates $K^+\rightarrow\pi^++\nu\bar\nu$.
There are numerous combinations of $\textit{c}_\mathcal{V}^{\,q}$,
$\textit{c}_\mathcal{V}^{\,l}$ and $d_\mathcal{U}$ which accord with
experiment.

\begin{figure}[h]
\raisebox{60pt}{\rotatebox{90}{\footnotesize $B\,(K^+ \rightarrow
\pi^+ +{\not}E)$}}
\includegraphics[width=300pt]{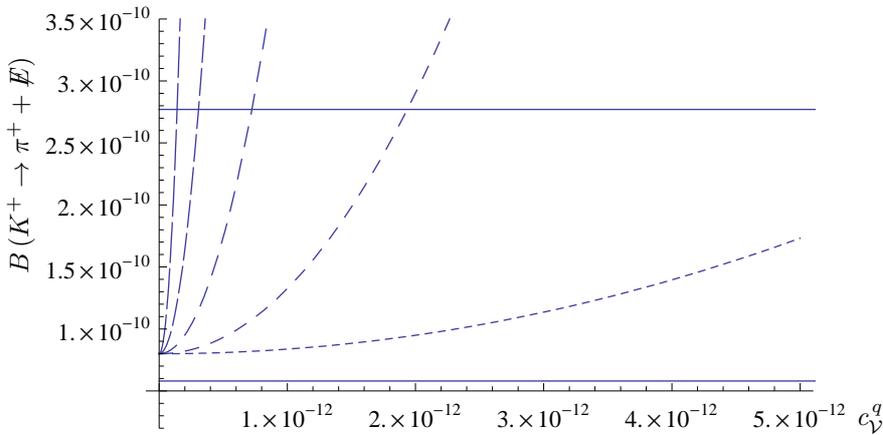}
\makebox(10,0)[br]{\footnotesize$\textit{c}_\mathcal{V}^{\,q}$}
\caption{\footnotesize Branching ratio versus
$\textit{c}_\mathcal{V}^{\,q}$ in the vector unparticle model with
$d_\mathcal{U}$=1.1, 1.3, 1.5, 1.7 and 1.9. The dash-length
increasess with $d_\mathcal{U}$. The unparticle mediation effect is
neglected. }
\end{figure}

When the couplings $c^l$'s are not very large, the mediation effects
are negligible. We plot in Fig. 4 the dependence of the branching
ratio $K^+\rightarrow\pi^+{\not E}$ on the coupling
$\textit{c}_\mathcal{V}^{\,q}$. The constraint on the coupling get
stronger wihen $d_\mathcal{U}$ increases. We  get roughly
$\textit{c}_\mathcal{V}^{\,q}\leq 10^{-13}$MeV$^{-d_\mathcal{U}}$ in
order to fit the experiments.

If the couplings $c^l$'s are large, the SM and the unparticle
contributions interfere in $K^+ \rightarrow \pi^+ +\nu \bar{\nu}$.
The interference can be either constructive or destructive,
depending on the number $d_\mathcal{U}$ and the couplings. We plot
the dependence of the total branching ratio $K^+ \rightarrow \pi^+
+{\not}E$ on $d_\mathcal{U}$ in Fig. 5 for
$\textit{c}_\mathcal{V}^{\,l}=-0.01$, and in Fig. 6 for
$\textit{c}_\mathcal{V}^{\,l}=0.01$. The branching ratio becomes
extremely large when $d_\mathcal{U}$ approaching 2 or 3. When
$d_\mathcal{U}$ approaches 2 or 3 the couplings
$\textit{c}_\mathcal{V}^{\,q}$ and $\textit{c}_\mathcal{V}^{\,l}$
are constrained strongly.

\begin{figure}[h]
\raisebox{60pt}{\rotatebox{90}{\footnotesize $B\,(K^+ \rightarrow
\pi^+ +{\not}E)$}}
\includegraphics[width=280pt]{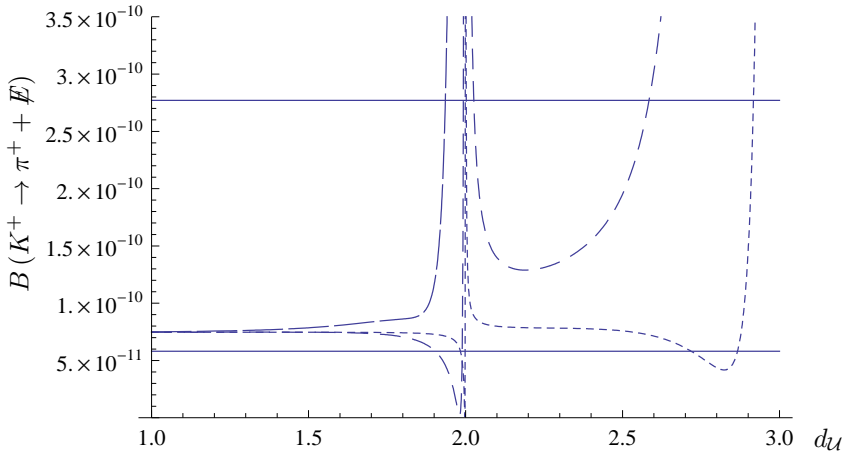}
\makebox(15,0)[br]{\footnotesize$d_\mathcal{U}$}
\caption{\footnotesize Branching ratio versus $d_\mathcal{U}$ in the
vector unparticle model with $\textit{c}_\mathcal{V}^{\,l}=-0.01$.
Here $\textit{c}_\mathcal{V}^{\,q}=1\times 10^{-15}$ for the
shortest dash-length, $\textit{c}_\mathcal{V}^{\,q}=4\times
10^{-14}$ for the middle dash-length, and
$\textit{c}_\mathcal{V}^{\,q}=1\times 10^{-13}$ for the longest
dash-length. }
\end{figure}

\begin{figure}[h]
\raisebox{60pt}{\rotatebox{90}{\footnotesize $B\,(K^+ \rightarrow
\pi^+ +{\not}E)$}}
\includegraphics[width=280pt]{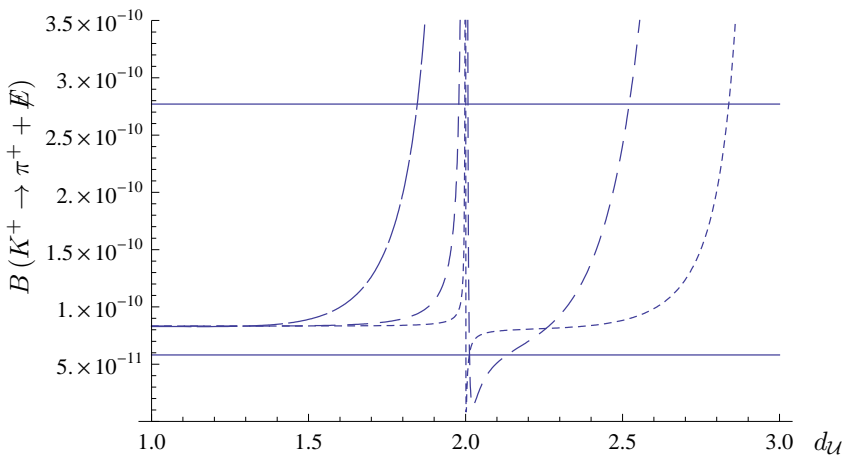}
\makebox(15,0)[br]{\footnotesize$d_\mathcal{U}$}
\caption{\footnotesize Same as in Fig. 5 except
$\textit{c}_\mathcal{V}^{\,l}=0.01$. }
\end{figure}

In Fig. 7 and Fig. 8 we show the dependence of the branching ratio
on the coupling $\textit{c}_\mathcal{V}^{\,q}$ for a very large
value of $\textit{c}_\mathcal{V}^{\,l}$ ($\mp -0.05$). We can find
that the vector unparticle contribution can be dominant in both
destructive and constructive cases, if $d_\mathcal{U}$ is large
enough.

\begin{figure}[h]
\raisebox{60pt}{\rotatebox{90}{\footnotesize $B\,(K^+ \rightarrow
\pi^+ +{\not}E)$}}
\includegraphics[width=280pt]{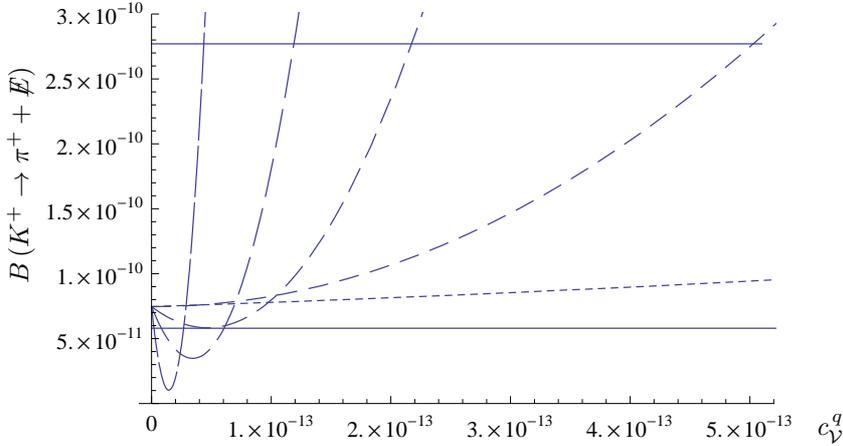}
\makebox(15,0)[br]{\footnotesize$\textit{c}_\mathcal{V}^{\,q}$}
\caption{\footnotesize Branching ratio versus
$\textit{c}_\mathcal{V}^{\,q}$ in the vector unparticle model with
$\textit{c}_\mathcal{V}^{\,l}=-0.05$ and $d_\mathcal{U}$=1.1, 1.5,
1.7, 1.8 and 1.9. The dash-length increase with $d_\mathcal{U}$.}
\end{figure}

\begin{figure}[h]
\raisebox{60pt}{\rotatebox{90}{\footnotesize $B\,(K^+ \rightarrow
\pi^+ +{\not}E)$}}
\includegraphics[width=280pt]{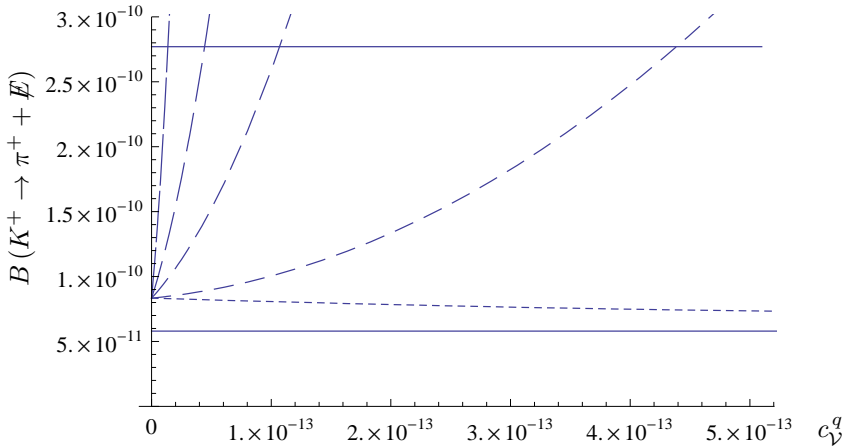}
\makebox(15,0)[br]{\footnotesize$\textit{c}_\mathcal{V}^{\,q}$}
\caption{\footnotesize Same as in Fig. 7 except
$\textit{c}_\mathcal{V}^{\,l}=0.05$.}
\end{figure}

The energy spectra for the charged $\pi$ are depicted in Fig. 9 and
Fig. 10 for the negative and positive values of
$\textit{c}_\mathcal{V}^{\,l}$'s, respectively. We find that for
large $\textit{c}_\mathcal{V}^{\,l}$'s, the spectra can be quite
different from the SM spectrum in the region when the $pi$'s are
hard. the spectrum is even much different from that in the scalar
unparticle model, due to the much complicated mediation of the
vector unparticle. If the difference in spectrum were found in the
future experiments, it would be clear signature as evidence of the
vector unparicle model.

\begin{figure}[h]
\makebox(20,160)[tr]{$\frac{d\Gamma^{{\not}E}}{dE_\pi}$}
\includegraphics[width=280pt]{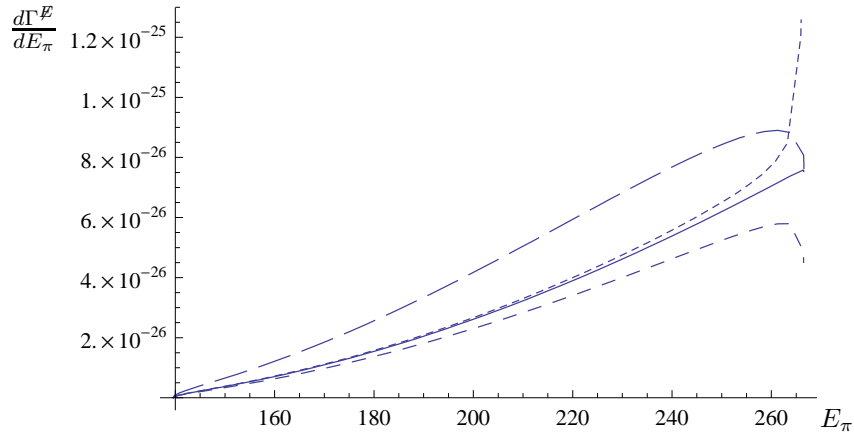}
\makebox(10,0)[br]{\footnotesize$E_\pi$} \par \caption
{\footnotesize {The energy spectra for the charged $\pi$ in the
vector unparticle model.  $d_\mathcal{U}$ = 1.3,
$\textit{c}_\mathcal{V}^{\,q} = 6.6\times10^{-14}$,
$\textit{c}_\mathcal{V}^{\,l} = -0.05$ for the shortest dash-length,
$d_\mathcal{U}$ = 1.8, $\textit{c}_\mathcal{V}^{\,q} =
5.0\times10^{-15}$, $\textit{c}_\mathcal{V}^{\,l} = -0.05$ for the
middle dash-length, $d_\mathcal{U}$ = 2.3,
$\textit{c}_\mathcal{V}^{\,q} = 1.0\times10^{-15}$
$\textit{c}_\mathcal{V}^{\,l} = -0.08$ for the longest dash-length.
The solid line represents the SM spectrum.}}\label{f}
\end{figure}

\begin{figure}[h]
\makebox(20,160)[tr]{$\frac{d\Gamma^{{\not}E}}{dE_\pi}$}
\includegraphics[width=280pt]{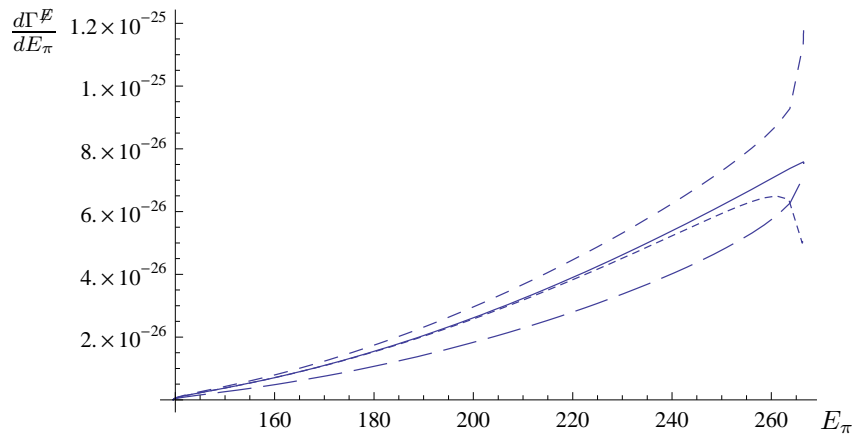}
\makebox(10,0)[br]{\footnotesize$E_\pi$} \par \caption
{\footnotesize {Same as in Fig.\ref{f} except
$\textit{c}_\mathcal{V}^{\,l}$'s are positive. }}
\end{figure}

\section{Conclusion}

In this paper we have discussed the process $K^+ \rightarrow \pi^+
+$Missing Energy in the unparticle models. We find that both the
branching ratio and the spectrum can be very different from the SM
predictions. Especially an vector unparticle model can bring some
complicated interference into the amplitude with the SM one.

This work was supported in part by the National Natural Science
Foundation of China (NSFC) under the grant No. 10435040.

\end{document}